\documentclass[amsmath,amssymb,prb,twocolumn,amssymb]{revtex4}

\usepackage{graphicx}
\usepackage{color}

\begin{document}
\title{Mesoscopic fluctuations of conductance of a helical edge contaminated by
magnetic impurities} 
\author{Vadim Cheianov}
\affiliation{Department of Physics, Lancaster University, Lancaster LA1 4YB, 
United Kingdom}
\author{Leonid I. Glazman}
\affiliation{Department of Physics, Yale University, New Haven, CT 06520, USA}
\date{\today}
\begin{abstract}
Elastic backscattering of electrons moving along the helical edge is prohibited
by time-reversal symmetry (TRS). We demonstrate, however, that an ensemble of
magnetic impurities may cause TRS-preserving quasi-elastic backscattering,
resulting in interference effects in the conductance. The characteristic energy
transferred in a backscattering event is suppressed due to 
the RKKY interaction of localized spins (the suppression is exponential in the
total number of magnetic impurities). We predict the statistics of conductance
fluctuations to differ from those in the conventional case of a one-dimensional
system with quenched disorder.
\end{abstract}
\maketitle
The constitutive porperty of a two-dimensional topological insulator is the presence
of helical edge states at its bounaries. The electron states propagating in
opposite directions along an edge form Kramers doublets and are protected
against elastic backsacattering by the time-reversal symmetry (TRS). The
fundamental consequence is the universality of the zero-temperature conductance
of a topological insulator. Experimental demonstration of such universality is
viewed as the confirmation of the existence of topological insulators. The
existing experiments clearly distinguish the topological insulators from
highly-resistive ``conventional'' ones~\cite{molenk2007}. However, the measured
conductance approaches the universal value only in very short (less than $1
\mu$m long) samples. Conductance of longer samples typically is lower,
indicating the presence of electron backscattering. 

Mechanisms of the backsacttering are a matter of ongoing debate. 
Current proposals utilize the Coulomb interaction between the
electrons of helical edge~\cite{kane-mele,moore,schmidt} or scattering of an
electron off a localized magnetic impurity~\cite{macejko1,macejko2,tanaka}. 
In all theoretical models considered so far the electron backscattering 
is either deeply inelastic or, as in the case of a magnetic impurity at 
temperatures exceeding the Kondo scale, quasi-elastic and incoherent 
due to the flips of the impurity spin~\cite{macejko1,tanaka}. 
In either case the bacskcattering, while suppressing the value of 
conductance $G$ at finite temperatures does not lead to mesoscopic 
fluctuations in $G$, in contrast with the low-temperature electron 
transport in conventional low-dimensional conductors. 

In this work we show that interference effects in the conductance of a helical
edge contaminated by an ensemble of magnetic impurities are actually possible.
Unlike a single impurity, spins in an ensemble are coupled by a long-range RKKY
interaction. The latter prevents individual spins from flipping in the course of
electron backscattering. That allows for a coherent superposition of the
electron de Broglie waves reflected by different impurities. Remarkably, this
picture does not contradict TRS: the electron reflection remains inelastic.
However, it is associated with a collective flip of a block of spins.
Therefore its amplitude is sensitive to the spatial structure of the electron
wave. Large number of spins in the block also ensures a parametrically small
(exponential in the number of spins) energy transfer in the scattering event.
This energy sets a mild lower limit for the temperature at which the considered
quantum-coherent phenomenon can be observed.

Interference in electron reflection manifests itself in mesoscopic
fluctuations of conductance $G,$  seen, for example 
when the chemical potential of electrons is continuously changed 
by a gate voltage $V_g$. 
In conventional one-dimensional wires such fluctuations are due to the
varying interference conditions for the potential scattering of electrons.
At sufficiently low temperatures such conductance fluctuations 
are pronouncedly non-Gaussian. In the regime of weak backscattering they obey 
the Rayleigh statistics. In contrast, we find that the variations of 
$G$ with $V_g$ caused by an ensemble of magnetic impurities tend to be close 
to gaussian except for temperature close to the spin-glass 
crossover.

We consider a topological insulator with a simple helical edge~\cite{molenk2007,kane-mele} 
such  that the spin $\mathbf s$ of an electron occupying an edge state has a 
conserved component $s_z$  in some fixed direction $\hat z.$  A magnetic impurity 
in a vicinity of the edge 
will experience two important 
interactions: a local single-ion anisotropy induced by the bulk spin-orbit coupling and 
a local exchange coupling to the electrons of the edge. (The direct exchange between the impurity spins is negligible at low impurity density.)
We assume the anisotropy 
to be of easy-axis type with some anisotropy constant $K>0.$ The exchange coupling 
of the edge-state electrons to the impurity spin will generally be anisotropic 
and depend on the position of the impurity. 
The effective low-energy Hamiltonian describing the helical edge with $N$ magnetic impurities is
\begin{align}
{\cal H}&={\cal H}_0-\sum_{i=1}^N K
\left({\mathbf n}\cdot{\mathbf S^i}\right)^{\!2}
+ \hbar v \sum_{i=1}^N \kappa_{ab}^i S^i_a
 \tau_b(x_i) \,,
\label{calH}\\
{\cal H}_0&= \hbar v\int dx\psi^\dagger (x)(-i\tau_z\nabla)\psi(x)\,.
\label{H0} 
\end{align}
Here the two-component spinor fields $\psi (x)$ represent the smooth (on the scale
provided by the Fermi wave length $2\pi/k_F$) envelope of the electron operators,
$\boldsymbol \tau(x)= \psi^\dagger (x)e^{-ik_F\tau_zx} \boldsymbol \tau e^{ik_F\tau_z x}\psi(x)$ is the 
electron spin density operator where $\boldsymbol \tau$ is the spin vector composed 
of the three Pauli matrices; 
$v$ is the electron velocity, $\mathbf S^i$ is the $i$'th impurity spin. We will see that
the interference effects in $G$ appear if the spin anisotropy axis $\mathbf n$
is different from $\hat z$ and the impurities have spin $S>1.$
We assume the magnetic impurities to be distributed 
randomly along the sample length $L$ and at random distances from the edge, 
resulting in random positions $x_i$ and coupling constants $\kappa_{ab}^i.$
In general, the exchange tensors $\kappa_{ab}^i$ and 
the tensor of single-ion anisotropy should be considered as running coupling 
constants, depending on the choice of the bandwidth cutoff. The 
renormalization of the anisotropy is not infrared-divergent. 
Therefore, assuming that the bare constants $\kappa_{ab}^i$ are
small, one adds an ultraviolet correction to Eq.~(\ref{calH}) of the form
\begin{equation}
\mathcal  H_{A} =\sum_{i=1}^N \delta K_{ab}^iS^i_aS^i_b
\label{HA}
\end{equation}
with $\vert \delta K^i \vert \sim \vert \kappa^i\vert^2 \Delta, $ where $\Delta$ is the bandwidth 
cutoff scale set by the insulator band gap. 
Renormalization of the tensors $\hat \kappa^i$ is
infrared-divergent, but remains small at energies exceeding the Kondo scale
$T_K$. The latter is exponentially small in $1/\vert\kappa^i \vert$ or even a higher 
power of that parameter, due to the effect of single-ion anisotropy. We note that the itinerant 
electrons facilitate the RKKY interaction between the impurities, which is of the order of 
$\hbar v \vert\kappa^1\vert \vert \kappa^2\vert /x$ for two impurities at distance $x$ from each other. 
We may ignore the renormalization of $\kappa$ as long as the RKKY exchange 
is larger than $T_K.$ 

We start the analysis of the model with considering two impurities at 
distance $x$ from each other. Treating the exchange tensors $\hat \kappa^i$ as 
perturbation theory parameters we find the leading-order RKKY interaction 
\begin{equation}
\label{RKKY}
\mathcal  H_{\rm RKKY} = - \frac{\hbar v }{4 \pi \vert x\vert}
S^1_{a}\kappa_{ab}^1\omega_{bc}(x) {P_{cd}} \kappa_{de}^2 S^2_e\,.
\end{equation}
Here $\omega_{bc}(x)$ is the orthogonal matrix of counterclockwise rotation 
through angle $2k_Fx$ about the $z$ axis
and ${ P_{cd}}=\delta_{cd}-\delta_{cz}\delta_{dz}$ is the matrix of 
orthogonal projection onto the $xy$ plane.
Assuming that  $\mathcal H_{\rm RKKY},$ Eq. \eqref{RKKY}, and $\mathcal H_{A},$ Eq. \eqref{HA}, 
are small as compared to the easy-axis anisotropy  $K$, we may apply secular
perturbation theory to determine the low-energy spectrum of the two-spin system. 
In the zeroth-order perturbation 
theory the ground state of the two-spin system is four-fold degenerate with the corresponding eigenspace 
spanned by four vectors, $|\pm S\rangle_1\otimes|\pm S \rangle_2$, where $\vert s \rangle_i$ 
denotes an eigenstate of ${\mathbf S}^i\cdot {\mathbf n}$ with an eigenvalue 
$s\in \{-S, -S+1, \dots , S-1, S\}$ and the subscript $i=1,2$ labels the Hilbert space 
attached to the $i$th spin.
The secular matrix of perturbation Eq.~(\ref{RKKY}) is conveniently written in terms of operators $\hat s_{i}$ 
acting on the $i$-th spin only:
\begin{equation}
\label{sigma-i}
\hat s_i=|S\rangle^{\phantom i}_i\langle S|^{\phantom i}_i-|-S\rangle^{\phantom i}_i\langle -S|^{\phantom i}_i\,.
\end{equation}
In this notation, the secular matrix takes form of Ising Hamiltonian,
\begin{equation}
\label{Ising1}
{\cal H}_I=-\delta E\cos(2k_Fx)\hat s_1 \hat s_2\,;
\end{equation}
here $\delta E=(\hbar v/4\pi |x|) n_{a}\kappa_{ab}^1 {P_{bc}}\kappa_{cd}^2 n_d.$ 
Depending on the sign of $\cos(2k_Fx)$ the ground state of the Hamiltonian (\ref{Ising}) is one of the two doublets,
 $|S\rangle_1 \otimes|S\rangle_2$, $|-S\rangle_1\otimes|-S\rangle_2$ or 
 $|S\rangle_1\otimes|-S\rangle_2$, $|-S\rangle_1\otimes|S\rangle_2$. 
The ground-state doublet is separated by energy $\sim|\delta E|$ from the excited level, which is also a doublet.
Note that for all $S>1$ the contribution of the perturbation Eq.~\eqref{HA} to the secular matrix is purely diagonal 
and has no effect on the splitting of the ground level.

In higher-order perturbation theory further splitting of the two doubly degenerate energy
levels occurs. The dominant effect here is due to the perturbation \eqref{HA}. Indeed, it 
has non-vanishing matrix elements for transitions in which the projection 
of one impurity spin is increased (or decreased) by $1$ or $2,$ 
for example $(\langle S-2|\!\otimes\!\langle S \vert) \mathcal H_A (\vert S\rangle\!\otimes\!|S\rangle)\neq 0.$
For $S>1$ any such transition takes a vacuum state to a virtual state 
having the energy of the order $K.$ At least $[S-1/2]$ such consecutive transitions (here 
the symbol $[\dots]$ stands for the integer part of $S$)
are needed in order to flip one impurity spin from $-S$ to $S,$
that is to bring the system from a vacuum state to a state in 
the excited doublet with energy $|\delta E|.$ Taking such processes into account 
amounts to introducing an off-diagonal correction to the Hamiltonian \eqref{Ising1}
\begin{equation}
\Delta\mathcal  H = \delta K \left ( \frac{\delta K}{K} \right)^{[S-1/2]}
\sum_{i=1,2}r_i\vert S \rangle_i \langle -S \vert_i\,,
\label{prt}
\end{equation}
where $r_{i}$ are some complex constants of the order of unity. If $\vert x\vert  \ll (\hbar v/\Delta)\times (K/\delta K)^{[S-1/2]}$, then each doublet 
in the spectrum of the Ising Hamiltonian~\eqref{Ising1} will split with the energy of the splitting
\begin{equation}
\label{epsilon}
\varepsilon \sim \delta K  
\left(\frac{\delta K}{K}\right)^{2[S-1/2]}\frac{\delta K}{|\delta E|}
\end{equation}
small compared to $|\delta E|$.

Consider now the combined dynamics of electrons and spins at energies $E\ll {\rm min}(\hbar v/x, K)$. 
It is described by the Hamiltonian ${\cal H}_{\rm eff}={\mathcal H}_0+{\mathcal H}_I+\Delta {\mathcal H}+{\mathcal U}$ with
\begin{equation}
\mathcal U=\hbar v \sum_{i=1,2} \hat{s}_i \psi^\dagger (x_i)
(\boldsymbol \xi_i \cdot \boldsymbol \tau)\psi(x_i)\,,
\label{U}
\end{equation}
where $\boldsymbol \xi_i = [P\omega(x_i) \kappa^i]\mathbf n$. We retain only the $x$ and $y$ components of $\boldsymbol\xi_i$ 
which cause electron backscattering,
\begin{equation}
\boldsymbol\xi_i=\xi_i\cdot (\cos 2k_Fx_i,\sin 2k_Fx_i,0)\,.
\label{xi}
\end{equation}
At temperatures $T\gg\varepsilon$ we may neglect the term $\Delta{\cal H}$ in ${\cal H}_{\rm eff}$. In that approximation, 
variables ${\hat s_i}$ are constants of motion. For each configuration of ${\hat s}_i$ we evaluate the backscattering current
within the Born approximation in ${\cal U}$. The total backscattering current is the Gibbs average of such contributions. 
It leads to the correction to the ballistic conductance,
\begin{equation}
\label{deltaG}
 \delta G = - \frac{e^2}{h} \left[\xi_1^2+\xi_2^2 + 2 (\boldsymbol \xi_1\cdot \boldsymbol \xi_2)\eta \right]\,.
\end{equation}
The term $2(\boldsymbol \xi_1\cdot \boldsymbol \xi_2)\eta $ here comes from the interference between the electron waves 
reflected by the two local magnetic moments. The factor $(\boldsymbol \xi_1\cdot \boldsymbol \xi_2)$ experiences the 
conventional Fabry-P\'erot oscillations as a function of the Fermi momentum $k_F$ with the period $2\pi/|x|$. The factor 
$\eta=\tanh [\delta E\cos(2k_Fx)/T]$ also oscillates with the same period; at low temperatures, $T\ll\delta E$, it
rapidly changes between $-1$ and $1$ each time $\cos 2k_Fx$ changes sign.

At the lowest energy scale, $T\lesssim\varepsilon$, one has to account for
$\Delta{\cal H}$. It lifts the degeneracy of the ground state and therefore prevents an electron with energy less 
than $\varepsilon$ from elastic backscattering within the helical edge. In order to investigate the electron 
conduction in this low-energy regime, we evaluate the backscattered current in a steady non-equilibrium state 
induced by the source-drain voltage. The application of the Fermi Golden Rule yields
\begin{multline}
I_{\rm BS}  = \frac{ve}{8\pi} \int_{-\infty}^\infty dp 
 \left[\xi_1^2+\xi_2^2+2 \eta(\boldsymbol \xi_1\cdot \boldsymbol \xi_2) \cos(2px) \right]
 \\
 \times 
\bigg \{
f_{++}(p)+f_{+-}(p)    - f_{-+}(p)- f_{--}(p)\\
+  
 \big[ f_{++}(p)f_{-+}(p) - f_{+-}(p)f_{--}(p) \big ]\tanh\frac{\varepsilon}{2T}
\bigg \}
\label{IBS}
\end{multline}
where 
\begin{equation}
 f_{\alpha\beta} (p)= \tanh\left(\frac{\hbar v p}{T}+ \alpha \frac{eV + \beta \varepsilon}{2T} \right), \quad \alpha, \beta=\pm.
\end{equation}
In the linear regime ($V\to 0$) Eq.~\eqref{IBS} predicts a correction to the 
ideal conductance 
\begin{equation}
\label{deltaG1}
 \delta G = -(\xi_1^2+\xi_2^2) F
 \left(
          {\textstyle \frac{\varepsilon}{T}, 0}
 \right) - 
 2\eta(\boldsymbol \xi_1\cdot \boldsymbol \xi_2)
 F
   \left(
     {\textstyle           \frac{\varepsilon}{T}, \frac{Tx}{\hbar v}  }
   \right) 
\end{equation}
with
\begin{equation}
F(w, z) = \frac{e^2}{h} \int_{-\infty}^{\infty} d \lambda \frac{2 \cosh ^2(\lambda ) \cos \left({2 z \lambda }
\right)}{\left[\cosh \left( w \right)+\cosh (2
   \lambda ) \right]^2}\,.
\end{equation}
The second argument of $F(w,z)$ is small in the entire region $T\ll\delta E/\kappa^2$, so we may set $z=0$. 
The $w\gg 1$ asymptote $F(w,0)\sim e^{-w}$ implies that the backscattering correction 
is exponentially suppressed at $T\ll \epsilon.$ In the opposite limit $w\to 0$ the function $F(w, 0)\to 1$ in 
agreement with Eq.~(\ref{deltaG}) obtained in the $\Delta{\cal H}=0$ approximation. Hereinafter we 
assume $T\gg\varepsilon$. 
\cite{footnote}

Next, we generalize the above considerations to a system of $N>2$ magnetic impurities statistically 
uniformly distributed with average density $n=N/L$ along the edge of length $L$. The corresponding effective 
Hamiltonian which allows one considering scattering of electrons with energies $E\lesssim \hbar v n$ has the 
form  ${\cal H}_{\rm eff}={\mathcal H}_0+{\mathcal H}_{\rm Ising}+\Delta {\mathcal H}+{\mathcal U}$,
where $\Delta {\mathcal H}$ and ${\mathcal U}$ are defined by Eqs.~(\ref{prt}) and (\ref{U}) 
(with extension of the summation to $N$), and
\begin{equation}
\mathcal H_{\rm Ising}=-\frac{\hbar v}{4\pi}
\sum_{i<j}\frac{\xi_i\xi_j\cos2k_F(x_i-x_j)}{|x_i-x_j|}\hat s_i \hat s_j\,,
\label{Ising}
\end{equation}
with $\xi_i$ defined in Eq.~(\ref{xi}).

The conductance correction evaluated in the Born approximation is
\begin{equation}
\delta G=-\frac{e^2}{h} \sum_{i,j}\xi_i\xi_j\cos2k_F(x_i-x_j)
\langle\hat s_i \hat s_j\rangle\,.
\label{deltaG2}
\end{equation}
Here the spin correlation function is
\begin{equation}
\langle\hat s_i \hat s_j\rangle=
\frac{{\rm Tr}\exp\left(-{\cal H}_{\rm Ising}/T\right)\hat s_i \hat s_j}
{{\rm Tr}\exp\left(-{\cal H}_{\rm Ising}/T\right)}\,.
\label{spinspin}
\end{equation}


In a given sample and at given temperature the conductance correction $\delta G,$ Eq.~\eqref{deltaG2}, depends on 
the Fermi momentum $k_F$ through the oscillatory factors in Eq.~\eqref{deltaG2} and in ${\cal H}_{\rm Ising}$, 
see Eq.~\eqref{Ising}. In experiment, this dependence take form of random fluctuations of the conductance 
as the Fermi energy of electrons at the edge is changed. The statistical properties of such fluctuations are 
fully encoded in the cumulants 
\begin{equation}
 G_m = \lim_{\lambda \to0} \frac{d^m}{d\lambda^m} \ln  \overline{\exp(\lambda \cdot \delta G)}
 \label{cumdef}
\end{equation}
where the overline represents the statistical average. 
We now demonstrate that the statistical properties 
of conductance fluctuations at the magnetically contaminated helical edge are 
drastically different from those in the case of the conventional quenched disorder in a
one-dimensional conductor. 

First, we recall the structure of conductance fluctuations caused by an ensemble of weak 
quenched scatterers in a usual one-dimensional conductor. At temperatures such that 
the coherence length and the thermal length $\hbar v/T$ are both greater than the sample 
length the conductance is temperature-independent. The electron reflection amplitude is, 
in the Born approximation, a linear superposition of $N$ random complex numbers, $e^{ik_F x_i},$
where $x_i$ is the position of the $i$th impurity. Consequently, in the large $N$ limit the 
conductance 
correction obeys the Rayleigh distribution, which is essentially non-Gaussian. 
In particular, for the second and third cumulants one has 
$
G_3= - 2 G_2^{3/2}.
$

In contrast, the fluctuations of conductance of the helical edge, Eq.~\eqref{deltaG2}, 
exhibit 
strong temperature dependence in 
the whole range of validity of the model \eqref{Ising}. We note that 
the model possesses an intrinsic temperature scale 
\begin{equation}
T_{\rm SG}= \frac{\hbar v\xi^2n}{4\pi}
\label{TSG}
\end{equation}
defining a crossover from the high-temperature regime, where spins thermally 
disordered to the ``spin glass'' regime where spins develop 
long-range correlations across the sample. For $T\gg T_{\rm SG}$ the model
can be investigated by means of the virial expansion. 

\begin{figure}
\includegraphics[width=0.48 \textwidth]{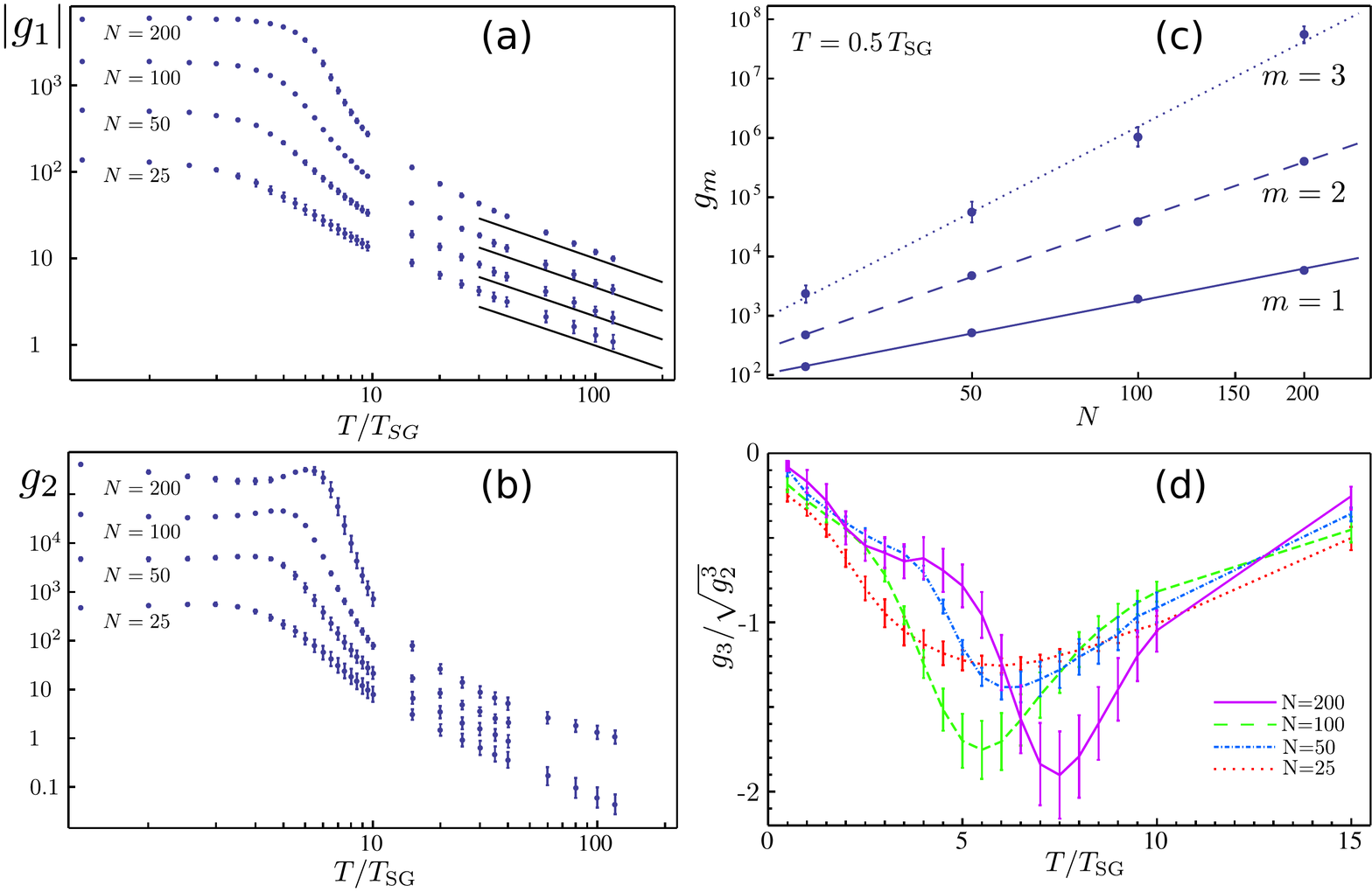}
\caption{MC data for the statistics of the mesoscopic conductance fluctuations at different temperatures
and system sizes. In panel (a) shown is the average conductance $g_1$ as a function of temperature. The 
asymptotes given by Eq.~\eqref{averG} are shown for comparison as solid black lines. 
Panels  (b) shows the second cumulant $g_2$ of the conductance distribution. 
Panel (c) shows the dependence of the first three cumulants at 
$T=0.5\, T_{\rm SG}$ on the system size. The log-log plot shows a good fit with the 
$g_m\sim N^{m+1}$ scaling. Panel (d) shows the skewness of the distribution of conductance 
values as a function of temperature.
}
\label{fig1}
\end{figure}

We assume that the coupling constants $\xi_i$ have a Gaussian distribution 
with the average $\xi$ and introduce the normalized cumulants $g_m$ such that
\begin{equation}
 G_m =\left(\frac{\xi^2 e^2}{h} \right)^m \left[g_m(\tau, N)- \delta_{m,1} N \right].
\end{equation}
The normalized cumulants are functions of the dimensionless temperature 
$\tau = T/T_{\rm SG}$ and the number of impurities $N=nL.$
At $T\gg T_{\rm SG}$ the spin-spin correlation function in 
Eq.~\eqref{deltaG2} is given by
\begin{equation}
\langle\hat s_i \hat s_j\rangle
=\tanh\left(\frac{\hbar v \xi_i\xi_j\cos 2k_F(x_i-x_j)}{4\pi |x_i-x_j| T}\right).
\label{spinspin1}
\end{equation}
Substituting Eq.~\eqref{spinspin1} in Eq.~\eqref{deltaG2} and averaging over the uniform 
distribution of magnetic impurities we find
\begin{equation}
g_1(\tau, N) =
-\frac{N}{8\pi \tau }\ln\left(\tau N \right) \,.
\label{averG}
\end{equation}
The logarithm in Eq.~\eqref{averG} is due to the $1/\vert x_i-x_j \vert$ dependence 
in the large-distance expansion of Eq.~\eqref{spinspin1}. The upper cutoff for this 
dependence is provided by the system size, $N/n,$ while the lower cutoff is defined by 
the distance $x$ at which the exchange energy $\hbar v \xi^2/x$ equals the 
temperature. For the higher cumulants the virial expansion yields
\begin{equation}
g_m(\tau, N) =  C_m\frac{ N}{\tau }, \qquad m>1.
\label{gm}
\end{equation}
Here $C_m$ are constants, in particular $C_2=0.20304$ and $C_3=0.01242.$ Note, that 
unlike the quenched case the third cumulant is positive. 
At $NT_{\rm SG}\gg T\gg T_{\rm SG}$  the higher cumulants 
satisfy $g_m^2\ll g_2^{m}$ therefore the distribution of conductance fluctuations is 
close to Gaussian.

\begin{figure}
\includegraphics[width=0.4 \textwidth]{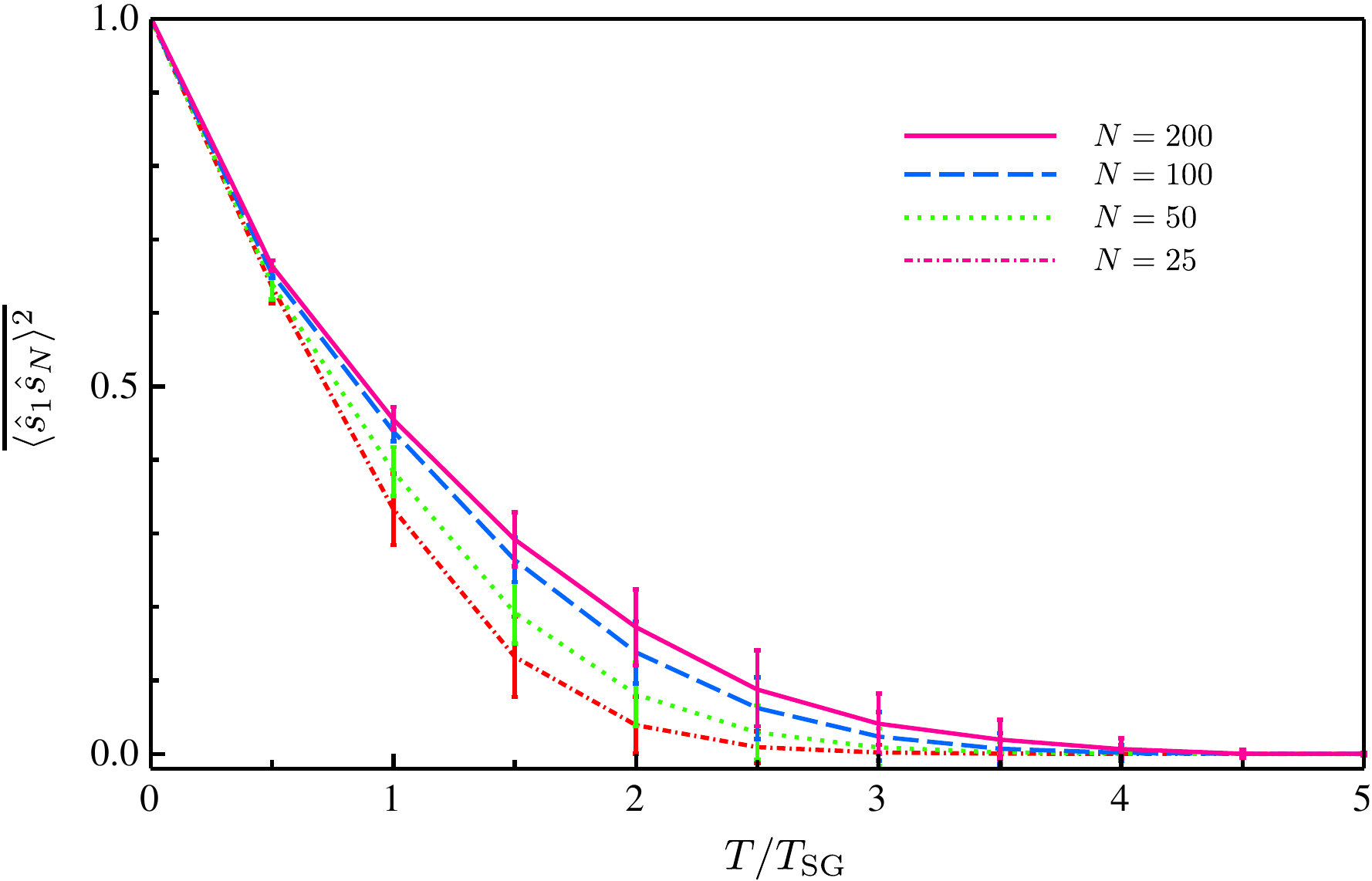}
\caption{Spin-spin correlation function at the opposite ends of the sample as a function of 
temperature for various systems sizes. Note that larger system sizes result in stronger 
correlations at given temperature. This effect is caused by the long-rangedness of 
the RKKY exchange resulting in the logarithmic renormalization of the spin-spin 
interaction constant with increasing system size.
}
\label{fig2}
\end{figure}

With decreasing temperature virial corrections to the spin-spin correlation function
\eqref{spinspin} become increasingly important. To explore this effect we calculate 
the first two terms in the virial expansion of the correlator
\begin{equation}
\overline{ \langle\hat{s}_i\hat{s}_j\rangle^2}=
\frac{1}{2\tau^2n^2|x_i-x_j|^2}
\left(1+\frac{2}{ \tau} \ln \tau n|x_i-x_j|+\dots \right)\, ,
\nonumber
\end{equation}
where $\dots$ stand for the higher-order terms in $1/\tau.$ One can see that no matter
how large the temperature $\tau$ is, the virial expansion breaks down  at sufficiently large distances. For a given system size 
$L$ one can define the crossover temperature $\tau_*$ such that
$1=\tau_*^{-1}\ln (n L\tau_*).$  Below this temperature equations \eqref{averG} and 
\eqref{gm} are not valid. The crossover temperature $\tau_*$ is 
a monotonically increasing function of the system size. In large systems 
there exists a parametric window $1<\tau<\tau_*$ where the long-range 
spin-glass correlations are absent, yet the virial expansion is invalid.

In order to investigate $g_m(\tau, N)$ at $\tau<\tau_*$ we performed Monte 
Carlo simulations for $N=25,50, 100$ and $200.$ For each $N$ we consider 
$10$ random realizations of quenched impurity positions, assuming $\xi$ to be 
the same for all impurities. We observe a considerable slowdown of the 
convergence of Metropolis algorithm for $T<5\, T_{\rm SG}$ caused by the 
onset of spin glass
correlations. To overcome this difficulty we employ parallel 
tempering, which works efficiently down to $T=0.5\, T_{\rm SG}.$ Numerical 
results for $g_{1,2}(\tau,N)$ 
are presented in Fig.~\ref{fig1}. In the temperature window 
$5\lesssim \tau \lesssim 10$ the cumulants experience a sharp increase with 
decreasing temperature from $g_m\sim N$ (see Eq.~\eqref{gm})
to $ g_m\sim N^{m+1}$ (see Fig.~\ref{fig1} (c) ).  
For $\tau \lesssim 5$ one can observe the onset of correlations  
between moments located near the opposite ends of the edge, Fig.~\ref{fig2}.
Note that at all temperatures the magnitude of the skewness of the conductance 
distribution is less than the universal value $2\sqrt 2$ predicted by 
the Rayleigh distribution. Moreover, at both high and low temperatures away 
from the ``spin glass'' transition the skewness is suppressed indicating a symmetric 
distribution of conductance, quite unlike the quenched case.


To conclude, the purpose of this work is to reconcile the possibility of mesoscopic 
fluctuations in the conductance of a helical edge with the absence of coherent backscattering in the 
presence of time-reversal symmetry (no external magnetic field applied). We find that scattering off an 
ensemble of large-spin ($S>1$) magnetic impurities may open a temperature window in which the conductance 
fluctuations are appreciable. The existence of such window is provided by a relatively strong effect of single-ion 
anisotropy which prevents easy flips of the impurity spins. It is further enhanced by the RKKY interaction between the
spins. The latter interaction depends on the Fermi momentum of helical edge, bringing ergodicity in the conductance 
fluctuations as a function of the helical edge chemical potential. 
We elucidated the signatures of the described mechanism in the  distribution function of conductance fluctuations.

This work was supported by Leverhulme Trust at Lancaster University; by the ERC grant No. 279738 – NEDFOQ 
and by NSF DMR Grant No. 1206612 at Yale University. 
LG acknowledges illuminating discussions with M. Goldstein and T.L. Schmidt.


\begin{thebibliography}{100}
\bibitem{footnote} Similar to $\delta G$, one may consider the differential conductance at $T=0$. The differential conductance 
correction $-\partial I_{\rm BS}/\partial V$ is exponentially small at $eV\ll\varepsilon$ and oscillates 
with $k_F$ at a fixed $eV\gtrsim\varepsilon$.
\end{thebibliography}
\end{document}